\documentclass[preprint,aps,prl,showpacs]{revtex4}
\usepackage{epsf, epsfig, graphics,amsmath}

\begin{document}
\draft
\title{Finite-size domains in membranes with active two-state
inclusions}
\author{Chien-Hsun Chen$^{(1),(2)}$ and Hsuan-Yi
Chen$^{(1),(2),(3)}$ } \affiliation{Department of
Physics$^{(1)}$,Center for Complex Systems$^{(2)}$, and Graduate
Institute of Biophysics$^{(3)}$,
National Central University, \\
Jhongli, 32054 \\
Taiwan}

\date{\today}

\begin{abstract}
The distribution of inclusion-rich domains in membranes with active
two-state inclusions is studied by simulations. Our study shows that
typical size of inclusion-rich domains ($L$) can be controlled by
inclusion activities in several ways. When there is effective
attraction between state-1 inclusions, we find: (i) Small domains
with only several inclusions are observed for inclusions with time
scales ($\sim 10^{-3} {\rm s}$) and interaction energy [$\sim
\mathcal{O}({\rm k_BT})$] comparable to motor proteins. (ii) $L$
scales as $1/3$ power of the lifetime of state-1 for a wide range of
parameters. (iii) $L$ shows a switch-like dependence on state-2
lifetime $k_{12}^{-1}$. That is, $L$ depends weakly on $k_{12}$ when
$k_{12} < k_{12}^*$ but increases rapidly with $k_{12}$ when $k_{12}
> k_{12}^*$, the crossover $k_{12}^*$ occurs when the diffusion
length of a typical state-2 inclusion within its lifetime is
comparable to $L$. (iv) Inclusion-curvature coupling provides
another length scale that competes with the effects of transition
rates.
\end{abstract}

\maketitle
 \section{Introduction}
Biological membranes are composed of lipids, proteins, carbohydrates
and other materials~\cite{ref:Lodish_book}. Many experimental as
well as theoretical studies have shown that the distribution of
these molecules in a membrane is nonuniform. Instead, dynamical
patch-like structures with typical size ranging from tens of
nanometers to about one micrometer exist in a
membrane~\cite{ref:Vereb_03}. The mechanism for the formation of
these heterogeneous structures has been a major research topic in
membrane biophysics in recent years. Possible mechanisms include (i)
continuous membrane recycling may result in nonequilibrium domains
in biological membranes~\cite{ref:recycling}, and (ii) the coupling
between lipid density and local membrane curvature provides a length
scale for membrane domains~\cite{ref:curvature}. The most important
conclusion from these studies is that finite-size domains in
membranes have to be sustained either by nonequilibrium processes
such as recycling or by effective long-range interactions due to
membrane elasticity.

Another important trend in recent biomembrane research is to treat
biomembranes as nonequilibrium
systems~\cite{ref:Rao_01,ref:Mouritsen_98,ref:Reigada_05-1,ref:Reigada_05-2,
ref:Prost_96,ref:Prost_00,ref:Prost_01,ref:Chen_04,ref:Lau_05} which
contain active inclusions that are driven by external stimuli.
Earlier work on these active membranes studied mainly the effect of
nonequilibrium activities on their fluctuation
spectrums~\cite{ref:Prost_96,ref:Prost_00,ref:Prost_01}, the
instabilities due to activities were also discussed
in~\cite{ref:Prost_00,ref:Rao_01}. In these models, the effect of
conformation change of the inclusions are neglected. On the other
hand, it has been shown by Lacoste and Lau~\cite{ref:Lau_05} that
active inclusions with an internal time scale have different
fluctuation power spectrum from simpler models which neglects this
effect. Furthermore, finite-wavelength instabilities were predicted
in~\cite{ref:Chen_04,ref:Reigada_05-1,ref:Reigada_05-2,ref:Mouritsen_98}
for membranes with two-state active inclusions, and it has been
pointed out in ~\cite{ref:Chen_04} that these instabilities are
similar to the dynamic nm-scale domains in biomembranes.
%  1.----
% Experimental relevance -- small clusters of BR in model membranes.
%
Indeed, recent experiments have observed enhanced spatial
aggregation of bacteriorhodopsin (BR) due to light-induced
activities in model membranes~\cite{ref:Kahya_02} and clustering of
light-harvesting antenna domains in photosynthetic membranes under
low-light conditions~\cite{ref:Scheuring_04}. These experimental
observations suggest that size of inclusion-rich domains in a
membrane can be tuned by changing the strength of external stimuli
of the active inclusions. Therefore in this article we study the
physics of activity-controlled finite size domains in membranes in a
simple model.

To focus on the basic interactions responsible for
inclusion-activity induced finite size membrane domains, we consider
the simplest possible model, i.e., a fluid membrane composed of one
type of lipid and two-state active inclusions. An inclusion is an
active unit which corresponds to a molecular complex that contains
proteins, lipids, and other molecules.  As shown schematically in
Fig.~1, inclusions in different internal states have different
conformations, therefore the interactions between the inclusions and
lipids depend on the internal states of the inclusions. There are
also couplings between the inclusion densities and membrane
curvature due to the up-down shape asymmetry of the inclusions.
Lattice Monte Carlo simulations are applied to study the
steady-state distribution of the inclusions when there is attractive
interaction between state-1 inclusions. The effect of
inclusion-curvature couplings and inclusion transition rates
$k_{\alpha \beta}$ (probability that a state-$\beta$ inclusion
transforms to state $\alpha$ per unit time) on the typical size of
inclusion-rich domains is discussed.
%  2.------
% Main results
%
The main results of our simulations are: (i) Small domains with only
several inclusions are observed for inclusions with time scales
($\sim 10^{-3} \ {\rm s}$) and interaction energy [$~\sim
\mathcal{O}({\rm k_BT})$] comparable to motor proteins, i.e., our
simple model is able to produce the kind of small inclusion clusters
observed in~\cite{ref:Kahya_02} within typical parameter range. (ii)
Typical size of inclusion-rich domain ($L$) scales as $L \sim
{k_{21}} ^{-1/3}$. This often observed
behavior~\cite{ref:recycling,ref:Reigada_05-1,ref:Reigada_05-2}
provides a mechanism to control continuously the size of
inclusion-rich domains by inclusion activities. (iii) A switch-like
response of typical size of inclusion-rich domains to the external
stimuli is found at fixed $k_{21}$: There exists a crossover
transition rate $k_{12}^*$ such that $L$ depends weakly on the
stimuli ($k_{12}$) when $k_{12} \lesssim k_{12}^*$ but becomes very
sensitive to it (i.e., $L$ increases rapidly as $k_{12}$ increases)
when $k_{12} \gtrsim k_{12}^*$.  The crossover $k_{12}^*$ occurs
when the diffusion length of a state-2 inclusion within its lifetime
is about the same as $L$.  (iv) $L$ decreases when the coupling
between state-1 inclusion and membrane curvature increases.

\section{Model} The system is described by a lattice model with
membrane height $h(i,j)$, inclusion densities $\phi_{\alpha}(i,j)$
($\alpha =1$ or $2$), and lipid density $\phi _0(i,j)$ defined on a
two-dimensional square lattice. The Hamiltonian of the system $H =
H_m + H_i + H_c$ includes the elastic energy of the membrane, the
interaction energy between the inclusions and lipid molecules, and
the inclusion-membrane coupling. To lowest order
\begin{eqnarray}
  H_m
= \frac{1}{2} \sum _{(i,j)} a^2 \left[ \kappa
 (\nabla_{\perp} ^2 h(i,j))^2 + \gamma ({\bf \nabla}_{\perp}h(i,j))^2
                                  \right] ,
\end{eqnarray}
where $\kappa$ is the bending rigidity, $\gamma$ is the surface
tension of the membrane, $a$ is the lattice constant, $\nabla
_{\perp}^2 h (i,j) = [
h(i+1,j)+h(i-1,j)+h(i,j+1)+h(i,j-1)-4h(i,j)]/a^2$, and $\nabla
_{\perp}h(i,j) = \hat{\bf i} [h(i+1,j)-h(i-1,j)]/2a + \hat{\bf j}
[h(i,j+1)-h(i,j-1)]/2a$.  $H_i$ includes the short-range
interactions between the inclusions and the lipids, the simplest
form is nearest-neighbor interaction
\begin{eqnarray}
 H_i =
 \sum _{\langle (i,j)(k,l) \rangle} \sum _{m,n=0}^2 J_{mn} \phi _m(i,j) \phi _n(k,l),
\end{eqnarray}
where $\sum _{m=0}^2 \phi _m(i,j)=1$ for all $(i,j)$, $\sum
_{\langle (i,j)(k,l) \rangle}$ means sum over all
nearest-neighboring pairs. The simplest form for the coupling
between the inclusion density and local membrane curvature is
\begin{eqnarray}
 H_c = \sum _{(i,j)} \sum _{m=0}^2 \kappa c_m \phi _m(i,j) [\nabla _{\perp} ^2
 h(i,j)],
\end{eqnarray}
where $c_0$ is the spontaneous curvature of the lipids, $c_{\alpha}$
($\alpha =1, \ 2$) is the coupling between state-$\alpha$ inclusions
and local membrane curvature. The inclusion-curvature coupling
constant $c_{\alpha}\sim \Delta \Sigma _{\alpha}/\Sigma _{\alpha}
l_{\alpha}$ has dimension of inverse length, where $\Sigma
_{\alpha}$ is the average cross sectional area of a type-$\alpha$
inclusion, $\Delta \Sigma _{\alpha}$ is the difference between head
area and $\Sigma _{\alpha}$, and $l_{\alpha}$ is the thickness (in
the $z$ direction) of the inclusion~\cite{ref:note_Safran}.

The simulations are performed on a $L_M \times L_M$ lattice with
periodic boundary condition. The lattice constant $a \sim 5$ nm is
chosen to be the smallest length scale for the continuum elasticity
theory of membranes to be valid~\cite{ref:lipowsky_PRL_99}. In the
simulations each lattice site is either occupied by a state-1
inclusion, a state-2 inclusion, or lipids. That is, $\phi _m(i,j)=
0$ or $1$ ($m=0, \ 1, \ {\rm or} \ 2$), and $\sum _{m=0} ^2\phi
_m(i,j)=1$ for all $(i,j)$. Since our goal is to study the effect of
$k_{\alpha \beta}$ and $c_m$ on the distribution of the
inclusion-rich domains, in the present study the active forces
exerted by the inclusions and the viscous flow of the solvent are,
similar to another active membrane simulation~\cite{ref:Weikl_06},
neglected.
%
%  This clarifies the simulation procedure:
%
Each Monte Carlo step is composed of three independent trial moves:
the motion of the membrane, the in-plane motion of the inclusions,
and the conformation change of the inclusions. The dynamics of
$h(i,j)$ is simulated by updating the membrane height of a randomly
chosen lattice site with a displacement between $-\Delta h$ and
$\Delta h$ with Metropolis algorithm, where $\Delta h = 0.4$ nm. The
in-plane motion of the inclusions is simulated by Kawasaki exchange
dynamics. Time interval of one Monte Carlo step can be derived from
the two-dimensional diffusion law $\langle r^2 \rangle = 4Dt$. Since
typical diffusion constant for membrane proteins is on the order of
$1 \mu {\rm m}^2/{\rm s}$~\cite{ref:Almeida_95}, and a free
inclusion moves a distance $a$ in one Monte Carlo step, a single
Monte Carlo step corresponds to time interval $\Delta t \sim 10^{-5}
{\rm s}$. The inclusion conformation changes are simulated by
assigning each state-$\beta$ (state-$\alpha$) inclusion a
probability $k_{\alpha \beta} \Delta t$ ($k_{\beta \alpha} \Delta
t$) to change its state to state $\alpha$ (state $\beta$) in each
Monte Carlo step. We choose $k_{\alpha \beta}$'s to be constants in
the simulations, i.e., we assume that the coupling between
conformation change of the inclusions and local membrane curvature
and composition is small. Therefore the conformation change does not
obey detailed balance, the steady state of the system is not in
thermal equilibrium~\cite{equilibrium}. %
In addition, in this article all the simulations are performed with
$k_{\alpha \beta} \Delta t < \mathcal{O}(1)$ such that the lifetime
of each state is sufficiently long that inclusion diffusion can be
taken into account properly. %
The initial condition of the simulations is randomly distributed
$N_{\rm inc}=\phi _{\rm inc} \times L_M \times L_M$ inclusions with
$N_{\rm inc}/2$ of them in state 1 and $N_{\rm inc}/2$ of them in
state 2 in a flat membrane. The probability that an arbitrary
inclusion is within an inclusion cluster of $M$ sites, $P(M)$, is
analyzed when the systems have reached steady states. Typical size
of inclusion-rich domain is taken as $L=\sqrt{M_{\rm max}}$, where
$M_{\rm max}$ is the value of $M$ which maximizes $P(M)$.

The free parameters in our model include $c_m$, $J_{mn}$ $(m=0, \ 1,
\ 2)$, and $k_{12}$, $k_{21}$.  $J_{mn}$ arises from short range
noncovalent interactions between the molecules including the effect
of hydrophobic length mismatch between different molecules or
inclusions of different internal states.  Since we are interested in
the case where finite-size inclusion-rich domains exist in the
steady state, $J_{mn}$ are chosen such that state-1 inclusions
attract each other, and state-2 inclusions have very weak or no
attraction with other inclusions and lipids.  To prevent aggregation
of state-2 inclusions due to membrane curvature-induced attraction,
$c_2$ has to be close to $c_0$. For state-1 inclusions, there is no
such restriction on the choice of $c_1$. The simplest choice of
parameters that satisfies the above criteria is $c_0=c_2=0$, $c_1
\geq 0$, $J_{00}=J_{11}=J_{22}=J_{02}=J$, and $J_{01}=J_{12}=J+
\Delta J$ with $\Delta J>0$, and this choice is applied to our
simulations. The bending elastic modulus and surface tension of the
membrane are chosen to be typical values, $\kappa = 5 \times
10^{-20}$ N$\cdot$m~\cite{ref:Strey_95}, and $\gamma = 24 \times
10^{-6}$ N$\cdot$m~\cite{ref:Needham_92}. $\Delta J \sim
\mathcal{O}({\rm k_BT})$ because the interactions between inclusions
and lipids are non-covalent, typically $c_1 \sim \Delta \Sigma
_{1}/\Sigma _{1} l_{1} \lesssim \mathcal{O}({1 \ \rm nm}^{-1})$. For
convenience we define a dimensionless constant $G=c_1 a
\sqrt{k_BT/\kappa}$ to describe the inclusion-curvature coupling,
its typical value is $G
\lesssim \mathcal{O} (1)$.%

The transition rates $k_{\alpha \beta}$ for the inclusions depend on
the mechanism of conformation transitions of the inclusions under
consideration.  (i) For inclusions have an intrinsic time scale
(like ion pumps), an inclusion stays at state $1$ (2) in the absence
of stimuli, and changes its conformation to state $2$ (1) when it
binds to energy source such as ATP or specific ions, the lifetime of
state $2$ (1) is an intrinsic property of the inclusion, it is on
the order of $10^{-2} - 10^{-3} \ {\rm s}$, therefore $k_{12} \
(k_{21}) \sim 10^2 - 10^3 \ {\rm s}^{-1}$. (ii) For inclusions that
change conformation by binding to or dissociating with small
ligands, the transition rates $k_{12}$, $k_{21}$ can exceed $10^3 \
{\rm s}^{-1}$, depending on the concentration of ligands in the
solvent.

\section{Discussion} Figure~2 shows domain size distribution $P(M)$
for $\phi_{\rm inc}=12.5\%$, $\Delta J=1.5 k_BT$, $k_{12}=10^{2} \
{\rm s}^{-1}$, $k_{21} = k_{12}/32$, and $G=0$, 1, and 2.  The peak
at $M=1$ comes from the isolated inclusions in the inclusion-poor
domain. The peak at greater $M$ provides the characteristic size of
inclusion-rich domains.  As $G$ increases, the characteristic size
of inclusion-rich domains decreases and the peak of $P(M)$ becomes
more significant due to inclusion-membrane coupling. This effect can
be better visualized by the snap shots from the steady states. As
shown in Fig.~3, when $G \neq 0$ the location of inclusion-rich
domains has strong correlation with the regions with high membrane
curvature. Since the system has lower free energy when the
inclusions are in the regions with higher curvature, thus the
membrane forms many mountain-like regions with inclusions in them.
This effect makes the width of $P(M)$ decrease as $G$ increases and
the position of the peak shifts to smaller $M$.

Figure~4 shows the relation between $L$ and $k_{21}$ for $\phi_{\rm
inc}=12.5\%$, $\Delta J=1.5 k_BT$, and $k_{12}=10^{-2} \ {\rm
s}^{-1}$ on a $128 \times 128$ lattice. %
The choice of $k_{12}$ is made to study inclusions with an intrinsic
time scale $\sim 10 \ {\rm ms}$, or inclusions in a solution with
high density ligands that induces $2 \rightarrow 1$ transitions. As
long as $\Delta J$ is order unity and $\phi _{\rm inc}$ is not too
small for the simulation system size, simulations with other choices
of $\Delta J$ and $\phi _{\rm inc}$ give results that are
qualitatively the same as Fig.~4. First of all, typical size of
inclusion-rich domains agrees with $L \sim {k_{21}}^{-1/3}$ pretty
well for a wide range of $k_{21}$, and the agreement is better when
$G$ is smaller~\cite{footnote}. This is because the growth of
inclusion-rich domains in the absence of inclusion activities is a
two-dimensional phase separation dynamics, i.e., $L \sim
t^{1/3}$~\cite{ref:Seul_94, ref:Bray_94, ref:Lipowsky_01}. This
growth eventually saturates due to state-1 to state-2 transitions on
time scale $\sim {k_{21}}^{-1}$, thus the typical length scales of
inclusion clusters in the steady state for $G \lesssim
\mathcal{O}(1)$ agrees with $L \sim
{k_{21}}^{-1/3}$~\cite{ref:Chen_04}. This relation is also observed
in simulations with other $\phi_{\rm inc}$, $J_{mn}$, and $k_{12}$
(data not shown). Finite size domains with
 $1/3$ scaling law due to similar mechanisms
have been observed in domains in membranes induced by continuous
recycling processes~\cite{ref:recycling}, and membranes made of
binary reactive lipids~\cite{ref:Reigada_05-1,ref:Reigada_05-2}.
Another interesting result is that small inclusion-rich clusters
with $M_{\rm max} \lesssim 10$ are observed when both  $k_{21}
\Delta t \gtrsim 10^{-3}$ (i.e, $k_{21} \gtrsim 10^2 \ {\rm
s}^{-1}$) and $k_{12} \Delta t \lesssim 10^{-2}$ (i.e., $k_{12}
\lesssim 10^3 \ {\rm s}^{-1}$) are satisfied.  Therefore in systems
with interaction energy and active transition rates comparable to
the characteristic time scale of a motor protein, the lifetime of
state-1 inclusions could be too short for a large cluster to be
formed. %
In ~\cite{ref:Kahya_02}, enhanced clustering of light-sensitive
bacteriorhodopsin (BR) is observed in the presence of light, but
there are only a few BR in a typical cluster.  A possible
explanation for this observation is that in the absence of light,
BRs have very weak interaction with each other (like our simulation
with $k_{12}=0$ and all inclusions are in state-2); while in the
presence of light, BRs in an intermediate state (analogous to
state-1 inclusions in our simulation) have attractive interaction
with each other, so BRs begin to form clusters. However, the
lifetime of this intermediate (for BR, it is on the order of $1$ ms)
is too short for large clusters to be formed, thus only very small
BR clusters are observed.

Figure~5 shows the relation between $L$ and $k_{12}$ for $\phi_{\rm
inc}=12.5\%$, $\Delta J=1.5 k_BT$, and $k_{21}
=10 \ {\rm s}^{-1}$ on a $128 \times 128 $ lattice. %
Here $L$ shows a switch-like dependence on $k_{12}$: $L$ depends
weakly on $k_{12}$ when $k_{12}$ is less than some crossover rate
$k_{12}^*$ but increases rapidly with $k_{12}$ when $k_{12} \gtrsim
k_{12}^*$. This can be understood by comparing $L_2\sim \sqrt{\left(
\frac{a^2}{\Delta t}\right) {k_{12}}^{-1}}$, the diffusion length of
a state-2 inclusion within its lifetime, to $L$. When $L_2 \gtrsim
L$, state-2 inclusions in inclusion-rich domains can escape to
inclusion-poor domains within their lifetime, thus $L$ depends
weakly on $k_{12}$. On the other hand, as illustrated in the inset
concentric circles of Fig.~5, when $L_2 \lesssim L$, the chance that
a state-2 inclusion in an inclusion-rich domain cannot escape to
inclusion-poor domains increases rapidly as $\Delta L = L - L_2$
increases.  In this case $L$ increases rapidly with $k_{12}$.  Thus
$k_{12}^*$ is determined by the condition $\sqrt{\left(
\frac{a^2}{\Delta t}\right) {{k_{12}^*}}^{-1}} \sim L$.  As can be
seen from Fig.~5, for given $G$, the crossover occurs when
$k_{12}\Delta t \sim (L/a)^{-2}$, and the switch-like behavior is
more significant for systems with smaller $G$.  For simulation with
$k_{21} > \mathcal{O}(10 \ {\rm s}^{-1})$, smaller $L$ causes
$k_{12}^*$ to be larger, and the switch-like behavior is less
significant because simulations are performed with $k_{12} \Delta t
\leq \mathcal{O}(1)$. In the limit of very large $k_{12}$, one
expects that almost no state-2 inclusion can escape from the domain
within its lifetime, and typical size of inclusion-rich domains
should saturate to its equilibrium value either determined by system
size or (when $G $ is large) by the length scale selected by
inclusion-membrane coupling. However, this limit happens when
$k_{12} \Delta t \geq \mathcal{O}(1)$, and therefore cannot be
accurately simulated in the present work.

The inclusion-curvature coupling $G$ also has important effects on
$L$. First, both Fig.~4 and Fig.~5 show that $L$ decreases as $G$
increases. That is, the coupling between inclusion density and
membrane curvature suppresses the formation of large inclusion
clusters because the inclusions tend to locate in regions with
greater membrane curvature. As a result, instead of forming large
inclusion-rich domains, the membrane prefers to form many
mountain-like regions with inclusions aggregating on them when $G
\neq 0$~\cite{ref:curvature}. Second, $L\sim {k_{21}}^{-1/3}$
relation in Fig.~4 does not describe large $G$ cases as well as
small $G$ cases because besides the length scale set by inclusion
transitions, inclusion-curvature coupling provides another length
scale for inclusion-rich domains. This is also true in Fig.~5, where
the rapid increase of $L$ at $k_{12}
> k_{12}^* \sim (L/a)^{-2}/\Delta t$ is not significant at large
$G$.  Thus, in general inclusion-curvature couplings competes with
the effects of $k_{\alpha \beta}$.

In summary, our study shows that there are several ways for
inclusion activities to control the typical size of nonequilibrium
membrane domains. In particular, the power-law relation $L \sim
{k_{21}}^{-1/3}$ provides a way to continuously tune the size of
domains, the switch-like dependence of $L$ on $k_{12}$ provides a
mechanism for sudden change of domain size dependence on the
inclusion activities. Although our model has neglected hydrodynamics
of the solvent and active forces exerted by the inclusions, the
mechanism proposed in the current study is rather general and should
exist in models which take these effects into account.  A detailed
analysis that includes the effects of active forces and full
hydrodynamics will be presented in a future work ~\cite{ref:future}.
The small inclusion clusters in the large $k_{21}$ regime observed
in our simulations are similar to the experimental observations
in~\cite{ref:Kahya_02}.  We expect more experiments on these systems
and other membranes can be performed to study the mechanisms
revealed in our model. For example, the different hydrophobic
lengths of different internal states of the
rhodopsin~\cite{ref:Isele_00} can be used to induce effective
interaction between nearby rhodopsin, therefore a
rhodopsin-containing membrane is also a good candidate for
experimental study.

 This work is supported by National Science Council of the Republic
of China under Grant No. NSC-93-2112-M-008-020.  Part of this work
is done during HYC's visit to the National Center for Theoretical
Sciences (NCTS), Hsinchu, Taiwan.

\vskip 1in

\newpage
\noindent
{\large Figure captions}\\
\begin{itemize}
\item FIG. 1: Schematics of a membrane with two-state
inclusions.

\item FIG. 2: Domain size distribution $P(M)$ for
$\phi_{\rm inc}=12.5\%$, $\Delta J=1.5 k_BT$, $k_{12}=10^{2} \ {\rm
s}^{-1}$, $k_{21} = k_{12}/32$, and $G=0$, 1, and 2.  The peak at
$M=1$ comes from the isolated inclusions in the inclusion-poor
domain. The peak at greater $M$ provides the characteristic size of
inclusion-rich domains.  As $G$ increases, the characteristic size
of inclusion-rich domains decreases and the peak of $P(M)$ becomes
more significant due to inclusion-membrane coupling.

\item FIG. 3: Snapshots for steady state inclusion distribution
(left, dark regions are occupied by inclusions) and membrane height
(right) for $\phi_{\rm inc}=12.5\%$, $\Delta J=1.5 k_BT$,
$k_{12}=10^{2} \ {\rm s}^{-1}$, and $k_{21} = k_{12}/32$. (a).
$G=0$, (b). $G=2$.  In order to faithfully present the morphology of
the membrane, the unit length for $h$ and the unit length in the
$xy$ plane are both chosen to be $a$.  For nonzero $G$ membrane
curvature close to the inclusion-rich domains increases and the
typical size of inclusion-rich domain decreases.

\item FIG. 4: Typical size of inclusion-rich domains in the
steady state for $\phi _{\rm inc} = 12.5 \%$, $\Delta J = 1.5 k_BT$,
$k_{12} \Delta t=10^{-3}$. $G=0$ (circles), $G=0.5$ (squares),
$G=1.0$ (diamonds), and $G=2.0$ (triangles).  The dashed line has
slope $-1/3$.

\item FIG.5: Typical size of inclusion-rich domains in the
steady state for $\phi _{\rm inc}=12.5\%$, $\Delta J = 1.5 k_BT$,
$k_{21} \Delta t= 10^{-4}$. $G=0$ (circles), $G=0.5$ (squares),
$G=1.0$ (diamonds), and $G=2.0$ (triangles).  Data with greatest
value of $k_{12} \Delta t$ are obtained from simulations on a 256
$\times$ 256 lattice.  The inset is a schematic of a domain with
radius $L$. When $k_{12}> k_{12}^*$, state-2 inclusions in the inner
circle (radius $\Delta L$) have small chance to leave the domain
within its lifetime. $\Delta L = L-L_2$ increases as $k_{12}$
increases.
\end{itemize}

\begin{figure}[h]
%\begin{left}
%\rotatebox{270}{
\epsfxsize= 4.5 in \epsfbox{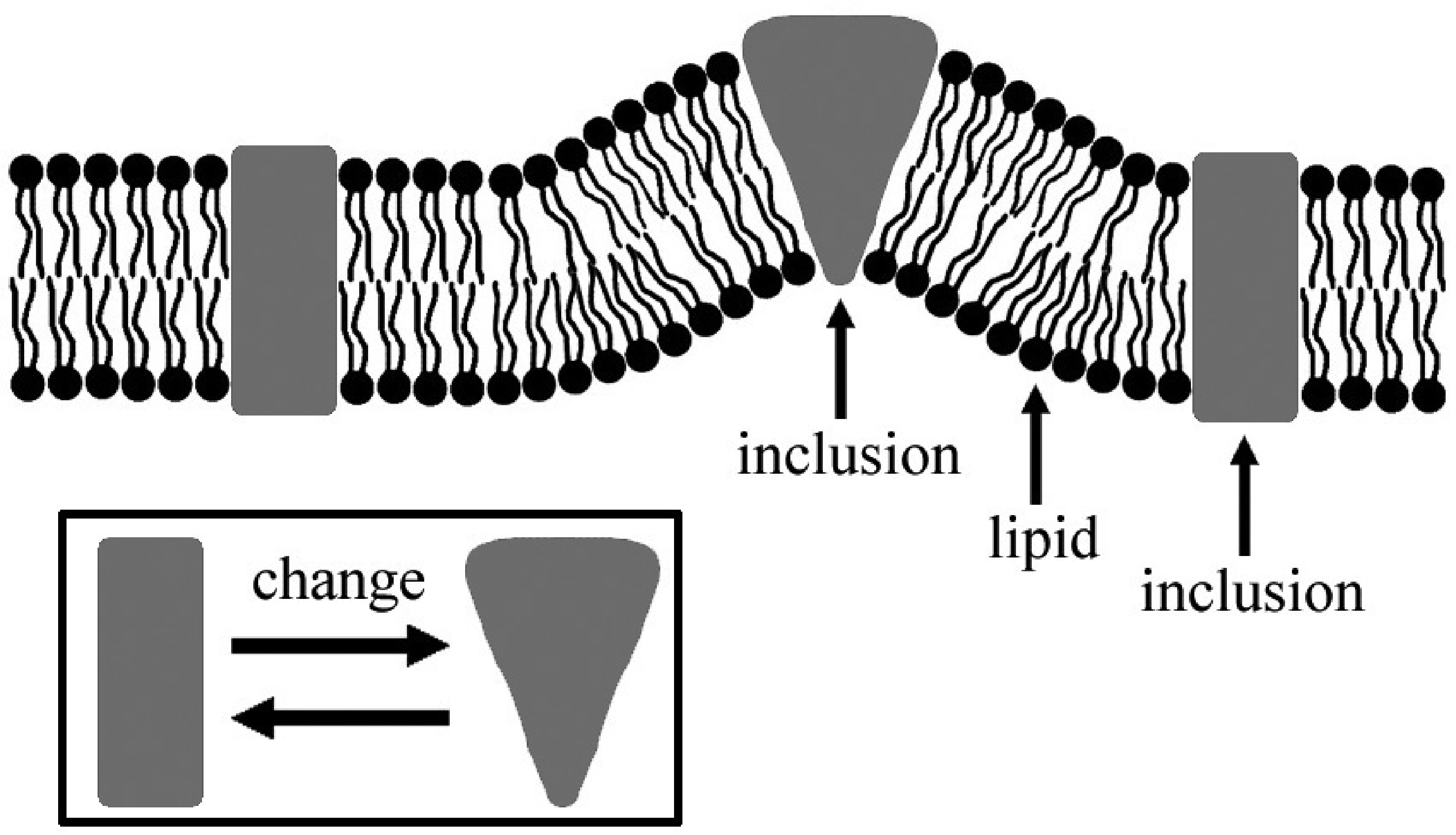}\caption{Schematics of a
membrane with two-state inclusions.}
%}
\end{figure}

\begin{figure}[h]
%\begin{left}
%\rotatebox{270}{
\epsfxsize= 4.5 in \epsfbox{fig2.eps}\caption{Domain size
distribution $P(M)$ for $\phi_{\rm inc}=12.5\%$, $\Delta J=1.5
k_BT$, $k_{12}=10^{2} \ {\rm s}^{-1}$, $k_{21} = k_{12}/32$, and
$G=0$, 1, and 2.  The peak at $M=1$ comes from the isolated
inclusions in the inclusion-poor domain. The peak at greater $M$
provides the characteristic size of inclusion-rich domains.  As $G$
increases, the characteristic size of inclusion-rich domains
decreases and the peak of $P(M)$ becomes more significant due to
inclusion-membrane coupling.}
%}
\end{figure}

\begin{figure}[h]
%\begin{left}
%\rotatebox{270}{
\epsfxsize= 4.5 in \epsfbox{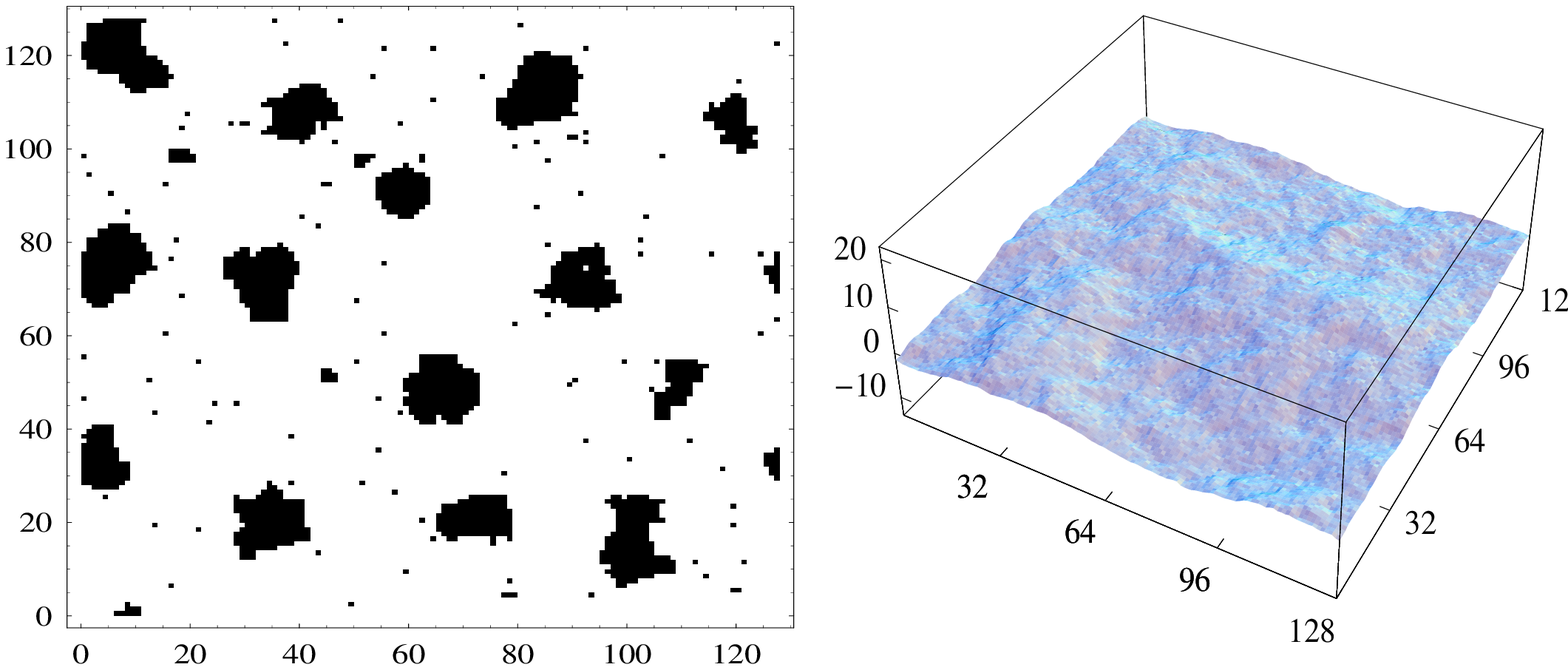} \epsfxsize=4.5 in
\epsfbox{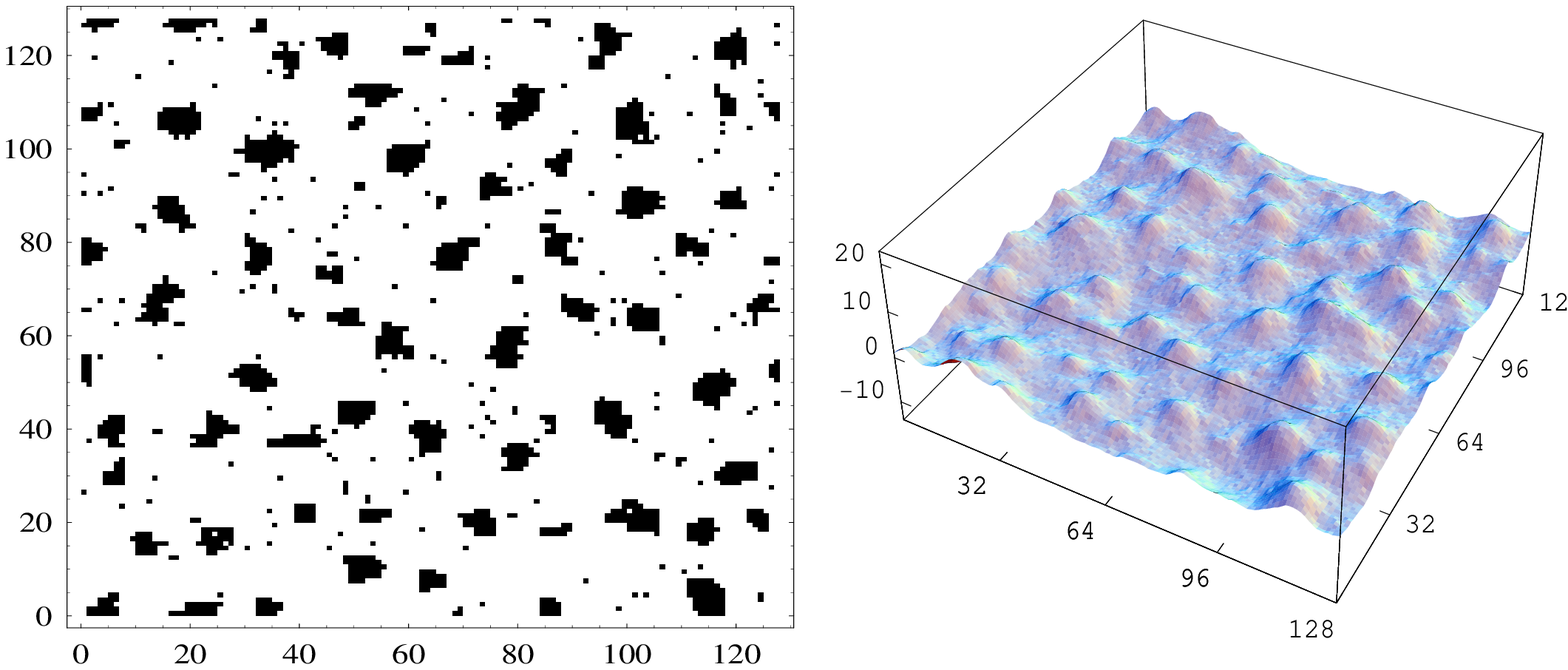}\caption{Snapshots for steady state inclusion
distribution (left, dark regions are occupied by inclusions) and
membrane height (right) for $\phi_{\rm inc}=12.5\%$, $\Delta J=1.5
k_BT$, $k_{12}=10^{2} \ {\rm s}^{-1}$, and $k_{21} = k_{12}/32$.
(a). $G=0$, (b). $G=2$.  In order to faithfully present the
morphology of the membrane, the unit length for $h$ and the unit
length in the $xy$ plane are both chosen to be $a$.  For nonzero $G$
membrane curvature close to the inclusion-rich domains increases and
the typical size of inclusion-rich domain decreases.}
%}
\end{figure}

\begin{figure}[h]
%\begin{left}
%\rotatebox{270}{
\epsfxsize= 4.5 in \epsfbox{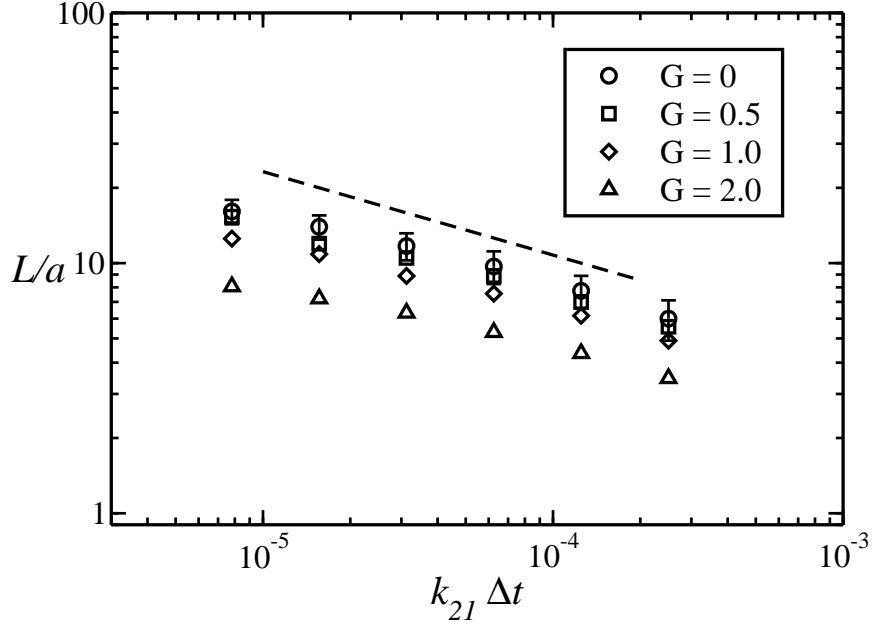}\caption{Typical size of
inclusion-rich domains in the steady state for $\phi _{\rm inc} =
12.5 \%$, $\Delta J = 1.5 k_BT$, $k_{12} \Delta t=10^{-3}$. $G=0$
(circles), $G=0.5$ (squares), $G=1.0$ (diamonds), and $G=2.0$
(triangles).  The dashed line has slope $-1/3$.}
%}
\end{figure}

\begin{figure}[h]
%\begin{left}
%\rotatebox{270}{
\epsfxsize= 4.5 in \epsfbox{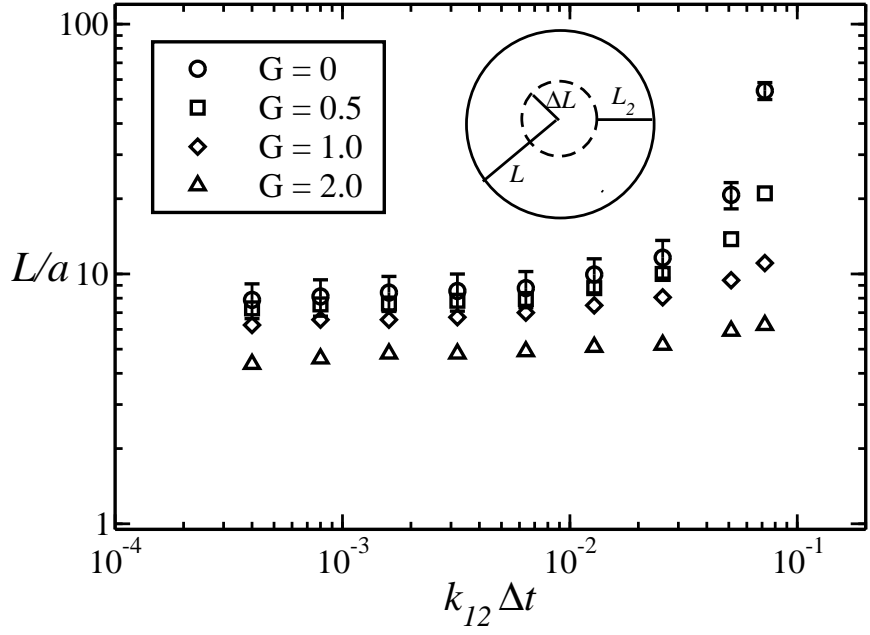}\caption{Typical size of
inclusion-rich domains in the steady state for $\phi _{\rm
inc}=12.5\%$, $\Delta J = 1.5 k_BT$, $k_{21} \Delta t= 10^{-4}$.
$G=0$ (circles), $G=0.5$ (squares), $G=1.0$ (diamonds), and $G=2.0$
(triangles).  Data with greatest value of $k_{12} \Delta t$ are
obtained from simulations on a 256 $\times$ 256 lattice.  The inset
is a schematic of a domain with radius $L$. When $k_{12}> k_{12}^*$,
state-2 inclusions in the inner circle (radius $\Delta L$) have
small chance to leave the domain within its lifetime. $\Delta L =
L-L_2$ increases as $k_{12}$ increases.}
%}
\end{figure}


\begin{references}
\bibitem{footnote} The small deviation of $L$ from the $-1/3$ power
law in the small $G$, small $k_{21} \Delta t$ region is due to
finite size effect because typical number of inclusions in a
cluster in our simulation becomes comparable to the total number
of inclusions in this region.
\end{references}
\end{document}